\pdfoutput=1

\documentclass[11pt]{article}

\usepackage[]{ACL2023}

\usepackage{times}
\usepackage{latexsym}
\usepackage{tipa}

\usepackage[T1]{fontenc}

\usepackage[utf8]{inputenc}

\usepackage{microtype}

\usepackage{inconsolata}
\usepackage{amsmath}
\usepackage{hyperref}
\usepackage{amssymb}
\usepackage{enumitem}
\usepackage{multirow}
\usepackage{xcolor}
\usepackage{color, colortbl}
\usepackage{tabularx}
\usepackage{caption}
\usepackage{booktabs}

\usepackage{arabtex}
\usepackage{utf8}
\setcode{utf8}
\usepackage{graphicx}
\usepackage{multirow}
\usepackage{amsmath,amssymb,amsfonts}
\usepackage{algorithmic}
\usepackage{url}
\usepackage{caption}
\usepackage{subcaption}

\title{Beyond Orthography: Automatic Recovery of Short Vowels \\ and Dialectal Sounds in Arabic}

\author{
  Yassine El Kheir\thanks{$^*$ These authors contributed equally to this work.}$^*$,
  Hamdy Mubarak,
  Ahmed Ali, 
  Shammur Absar Chowdhury$^*$\footnotemark[1]\thanks{ $^+$ Corresponding author. }$^+$ \\
  \{yelkheir, hmubarak, amali, shchowdhury\}@hbku.edu.qa
}

\begin{document}

\maketitle

\begin{figure} [ht]
\centering
\includegraphics[width=0.4\textwidth]{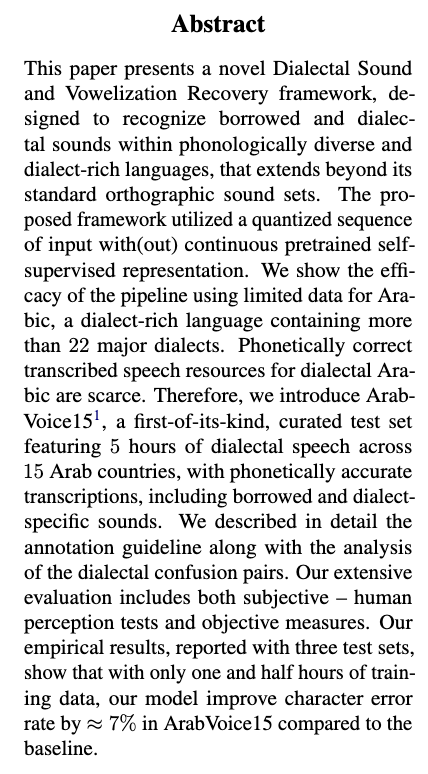}
\end{figure}


\section{Introduction}


Self-supervised learning (SSL) paradigm has transformed speech research and technology, achieving remarkable performance \cite{baevski2020wav2vec,chen2022wavlm} while reducing the dependency on extensively annotated datasets \cite{radford2023robust}. The SSL models excel at discerning the underlying acoustic properties in both frames and utterance level \cite{pasad2021layer, pasad2023comparative, chowdhury2023end} irrespective of language. Phonetic information is sailent and preserved even when these continuous representations are mapped to a finite set of codes via vector quantization \cite{Hsu2021, Sicherman2023,Wells2022,ElKheir2024}. This allows the learning paradigm to leverage unlabeled data to discover units that capture meaningful phonetic contrasts. 

Leveraging insights from acoustic unit discovery \cite{Park2008,Versteegh2015, Dunbar2017, eloff2019unsupervised, van2020vector}, unsupervised speech recognition \cite{baevski2021unsupervised,Liu2018,Chen2019,Liu2022,Baevski2021}, and phoneme segmentation \cite{Kreuk2020, Bhati2022,Dunbar2017, Versteegh2015} have utilized quantized discrete units for various purposes. 
These include \textit{(i)} pretraining the SSL model \cite{baevski2020wav2vec, Hsu2021},  \textit{(ii)} employing acoustic unit discovery as a training objective \cite{VanNiekerk2020}, and \textit{(iii)} utilizing discrete labels for training phoneme recognition and automatic speech recognition \cite{chang2023exploration,Liu2022,Liu2018,Sukhadia2024}. 

Inspired by previous research, we employ SSL representations and vector quantization to recognize acoustic units in phonologically diverse spoken dialects, extending beyond their standard orthographic sound sets. We introduce a simple yet potent network leveraging SSL and a discrete codebook to recognize these non-orthographic dialectal and borrowed sounds with minimal labeled data.


Arabic is an appropriate language choice for the task. 
The language has a rich tapestry of dialects, each with its unique characteristics in phonology, morphology, syntax, and lexicon \cite{ali2021connecting}. These dialects\footnote{There are $22$ Arab countries, and typically, there is more than one dialect spoken in each Arab country (ex: rural versus urban areas)} differ not only among themselves but also when compared to Modern Standard Arabic (MSA). While MSA prevails in official and educational domains, Dialectal Arabic (DA) serves as the means for daily communication. 
The diversity in pronunciation and phoneme sets for DA goes beyond standardized MSA sound sets. Moreover, to add to the challenges, DA follows no standard orthography. Therefore, despite the abundance of DA speech data in online platforms, accurately (phonetically correct) transcribed resources are scarce, categorizing DA among the low-resource languages. 

To bridge this gap, we introduce the Arabic ``\textit{Dialectal Sound and Vowelization Recovery}'' (DSVR) framework. The proposed framework exploits the frame-level SSL embeddings and quantizes them to create a handful of discrete labels using k-means model. These discrete labels are then fed (can be in combination with SSL embeddings) as input to a transformer-based dialectal unit and vowel recognition (DVR) model. 

We show its efficacy for \textit{(a)} dialectal and borrowed sound recovery; and \textit{(b)} vowelization restoration capabilities with only 1 hour 30 minutes of training data.
We introduced Arabic dialectal test set -- ``\textbf{ArabVoice15}'', a collection of $5$ hours of dialectal speech and verbatim transcription with recovered dialectal and borrowed sounds from 15 Arab countries. For vowelization restoration, we tested on $1$ hour of speech data, sampled from CommonVoice-Ar \cite{ardila2019common}, transcribed by restoring short vowels. Our paper describes the phonetic rules adopted, special sounds considered along with detailed annotation guidelines for designing these test sets. 
Furthermore, we evaluate the quality of the intermediate discrete labels using human perceptual evaluation, in addition to other purity and clustering-based measures. 

We observed that these discrete labels can capture speaker-invariant, distinct acoustic, and linguistic information while preserving the temporal information. Consequently, encapsulating the discriminate acoustic unit properties, which can be used to recover dialectal missing sounds. 
Our empirical results suggest that DSVR can exploit unlabeled data to design the codebook and then with a small amount of annotated data, a unit recognizer can be trained.

Our contribution involves: 
    \textit{(i)} Proposed Arabic Dialectal Sound and Vowelization Recovery (DSVR) framework to recognize dialectal units and restore short vowels;
    \textit{(ii)} Developed annotation guidelines for the verbatim dialectal transcription;
    \textit{(iii)} Introduced and benchmark ArabVoice15 test set -- a collection of dialectal speech and phonetically correct verbatim transcription of 5 hours of data.
    \textit{(iv)} Released a small subset of CommomVoice - Arabic \cite{ardila2019common} data with restored short vowels, dialectal and borrowed sounds.

\noindent This study addresses the crucial challenge of identifying and understanding these phonetic intricacies, acknowledging their essential role in improving the performance of speech processing applications like dialectal Text-to-Speech (TTS) and Computer-Assisted Pronunciation Training applications. To the best of our knowledge, this study is the first to attempt to automatically restore vowels, borrowed and dialectal sounds for rich spoken dialectal Arabic language with very limited amount of data. Moreover, the study also introduce the very first dialectal testset with phonetically correct transcription representation.

\label{sec:related}

\section{Arabic Sounds}
\label{sec:arabsounds}
The exploration of phonotactic variations across Arabic dialects, including MSA and other regional dialects offers a rich field of study within the domain of Arabic linguistics. These variations are not merely lexical, but phonetic and in many cases deeply embedded in the phonological rules that dictate the permissible combinations and sequences of sounds within each dialect \cite{biadsy2009spoken}.

\subsection{Related Studies}
Limited research investigated dialectal sounds in Arabic transcribed speech. 
\citet{vergyri2004automatic} deployed an EM algorithm to automatically optimize the optimal diacritic using acoustic and morphological information combination. \citet{al2014lexical} employed automated text-based diacritic restoration models to add diacritics to speech transcriptions and to train speech recognition systems with diacritics. However, the effectiveness of text-based diacritic restoration models for speech applications is questionable for several reasons, as demonstrated in \citet{aldarmaki2023diacritic}, 
they often fail to accurately capture the diacritics uttered by speakers due to the nature of speech; hesitation, unconventional grammar, and dialectal variations. 
This leads to a deviation from rule-based diacritics. Recently, \citet{shatnawi2023automatic} developed a joint text-speech model to incorporate the corresponding speech signal into the text based diacritization model. 

Grapheme to Phoneme (G2P) has been studied thoroughly by many researchers across multiple languages. Recent approaches in G2P include data-driven and multilingual \cite{yu2020multilingual,garg2024data} mapping from grapheme sequence to phoneme sequence. However, previous work in Arabic G2P is comprised of two steps: (\textit{i}) Grapheme to vowelized-grapheme (G2V)  to restore the missing short vowels and (\textit{ii})  Vowelized-grapheme to phoneme sequence (V2P). The first step is often statistical and deploys techniques like sequence-to-sequence; for example studies like \citet{abdelali2016farasa, obeid-etal-2020-camel} are used widely for restoring the missing vowels in Arabic. The second step is relatively one-to-one and can be potentially hand-crafted rules for MSA as well as various dialects, refer to \citet{biadsy2009spoken,ali2014complete} for more details. MSA Arabic speech recognition phoneme lexicon can be found here\footnote{\url{https://catalog.ldc.upenn.edu/LDC2017L01}}

The distinction between MSA and regional dialects is nuanced; viewing them as separate is oversimplified. Arabs perceive them as interconnected, leading to diglossia, where MSA is for formal contexts and dialects for informal ones, yet with significant overlap and blending \cite{chowdhury2020does}. 
\citet{chowdhury2020effects} studied dialectal code-switching in the manually annotated Egyptian corpus. 
The corpus was annotated for both MSA and Egyptian dialect labels per token, considering both the linguistic and the acoustic cues. The findings indicate the complex overlapping characteristics of the dialectal sound units showing roughly $2.6K$ Egyptian sounding words with respect to $9.3K$ MSA and 2.3K mix of both.  

\subsection{MSA and Dialectal Phonlological Variations}
Arabic dialects exhibit phonological differences when compared to MSA, these differences might be noted across various aspects of pronunciation and phonology, such as consonants, vowels, and diphthongs. It's suggested that Arabic generally encompasses around $28$ consonants, alongside three short vowels, three long vowels, though these numbers could vary slightly depending on the dialect in question.
The consonant pronunciation of 
\<ث>\ [\textipa{$\theta$}],
\<ذ>\ [\textipa{\dh}],
\<ظ>\ [\textipa{\dh\textsuperscript{Q}}], 
\<ج>\ [\textipa{d\textctz}],  \<ض>\ \textipa{[d\super Q], and \<ق>\ [\textipa{q}]} 
cover most of the variations across Arabic dialects. Here are some examples of phones that vary between MSA and various Arabic dialects. 
\begin{itemize}[noitemsep,topsep=0pt,parsep=0pt,partopsep=0pt]
\item Interdental Consonants: In particular 
\<ث> [\textipa{$\theta$}]/
\<ذ> [\textipa{\dh}]
found in MSA are pronounced differently. For example, in Egyptian Arabic, they are often pronounced as 
\<س> [\textipa{s}]. 

\item  The voiceless stop constant 
\<ق>\ [\textipa{q}] 
is a good example across Arabic dialects, In many cases, it will be pronounced as glottal stop 
\<ء>\ [\textipa{\textglotstop}] 
in Egyptian dialect and voiced velar
\<ج>\ [\textipa{d\textctz}] 
in Gulf and Yemeni dialects.
\item Long and short vowels might exhibit a reduction in duration or even drop in duration in various dialects. In some dialects, the difference between long and short vowels may be subtle to notice.  
\item The difference in stress between Arabic dialects can lead to different meanings.
\end{itemize}
The phonological differences and examples mentioned above do not cover all variations but highlight several distinctions between Arabic dialects and MSA. A depiction of certain MSA sound variations is presented in Appendix \ref{Appendix1}.

\section{Methodology}



\begin{figure*} [ht]
\centering
\includegraphics[width=1\textwidth]{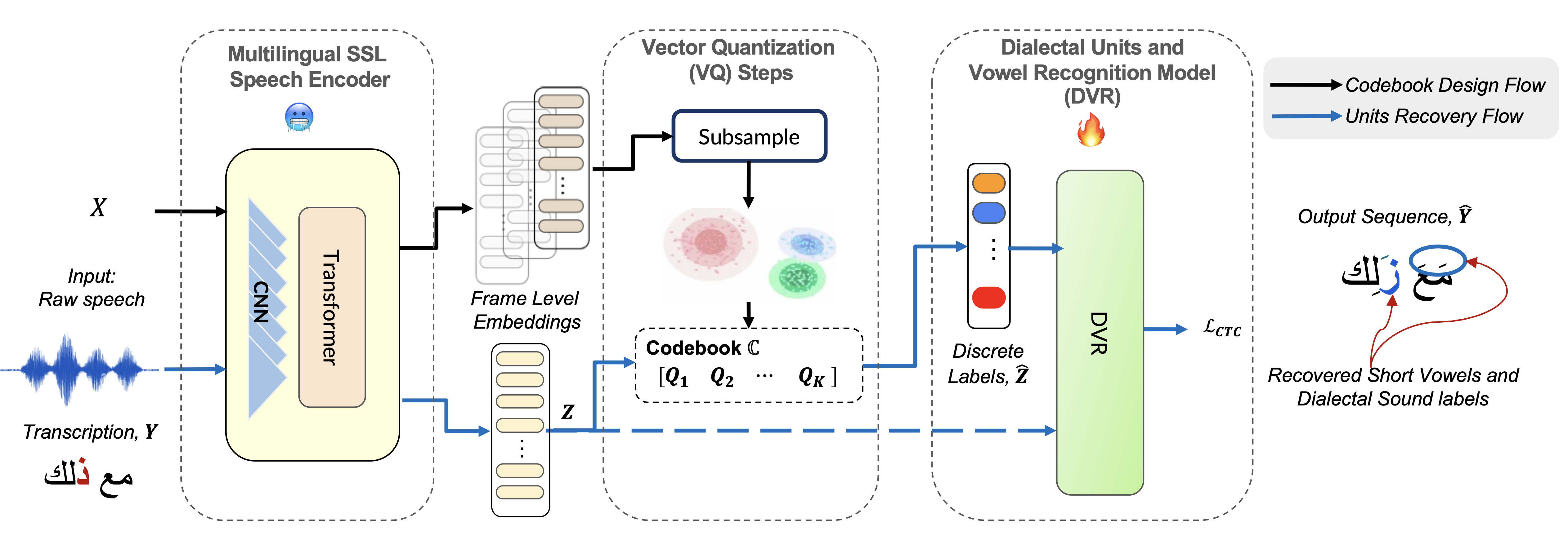}
\caption{\small{Proposed Arabic Dialectal Sound and Vowelization Recovery (DSVR) Framework}}
\label{fig:exp_flow}
\end{figure*}

Figure \ref{fig:exp_flow} gives an overview of our proposed \textit{Dialectal Sounds and Vowelization Restoration Framework}. The goal of the pipeline is to recover (verbatim) dialectal sound and short vowel units, using frame-level representation. Given an input speech signal $X=\left [ x_1, x_2, \cdots , x_T \right ]$ of T frames, the frame-level representation ($Z$) is first extracted from a \textit{multilingual SSL pretrained} model. 

We subsampled frame-level vectors ($\widetilde{Z} \subset Z$) to train a simple \textit{Vector Quantization} (VQ) model
using k-means for getting a Codebook $\mathbb{C}_{k}$, with $k$ categorical variables. Each cluster, in the codebook, is then associated with a code $Q_{i}^k$ and a centroid vector $G_{i}^k$. Using the $\mathbb{C}_{k}$ codebook, we infer the discrete sequences codes $\hat{Z}$ corresponding to the input $Z$. $\hat{Z}$ is the input of our \textit{Dialectal Units and Vowel Recognition} (DVR) module.



\subsection{Pretrained Speech Encoder}
\label{sec:xlsr}
The XLS-R\footnote{https://huggingface.co/facebook/wav2vec2-large-xlsr-53} model is a multilingual pre-trained SSL model following the same architecture as wav2vec2.0 \cite{baevski2020wav2vec}.
It includes a CNN-based encoder network to encode the raw audio sample and a transformer-based context network to build context representations over the entire latent speech representation. The encoder network consists of $7$ blocks of temporal convolution layers with $512$ channels, and the convolutions in each block have strides and kernel sizes that compress about $25$ms of $16$kHz audio every $20$ms. The context network consists of $24$ blocks with model dimension $1024$, inner dimension $4096$, and $16$ attention heads. 

The XLS-R model has been pre-trained on around $436,000$ hours of speech across $128$ languages. This diverse dataset includes parliamentary speech ($372,000$ hours in $23$ European languages), read speech from Multilingual Librispeech ($44,000$ hours in 8 European languages), Common Voice ($7,000$ hours in $60$ languages), YouTube speech from the VoxLingua107 corpus ($6,600$ hours in $107$ languages), and conversational telephone speech from the BABEL corpus ($\approx$ $1,000$ hours in $17$ African and Asian languages).

\begin{figure*} [ht]
\centering

\includegraphics[width=0.7\textwidth]{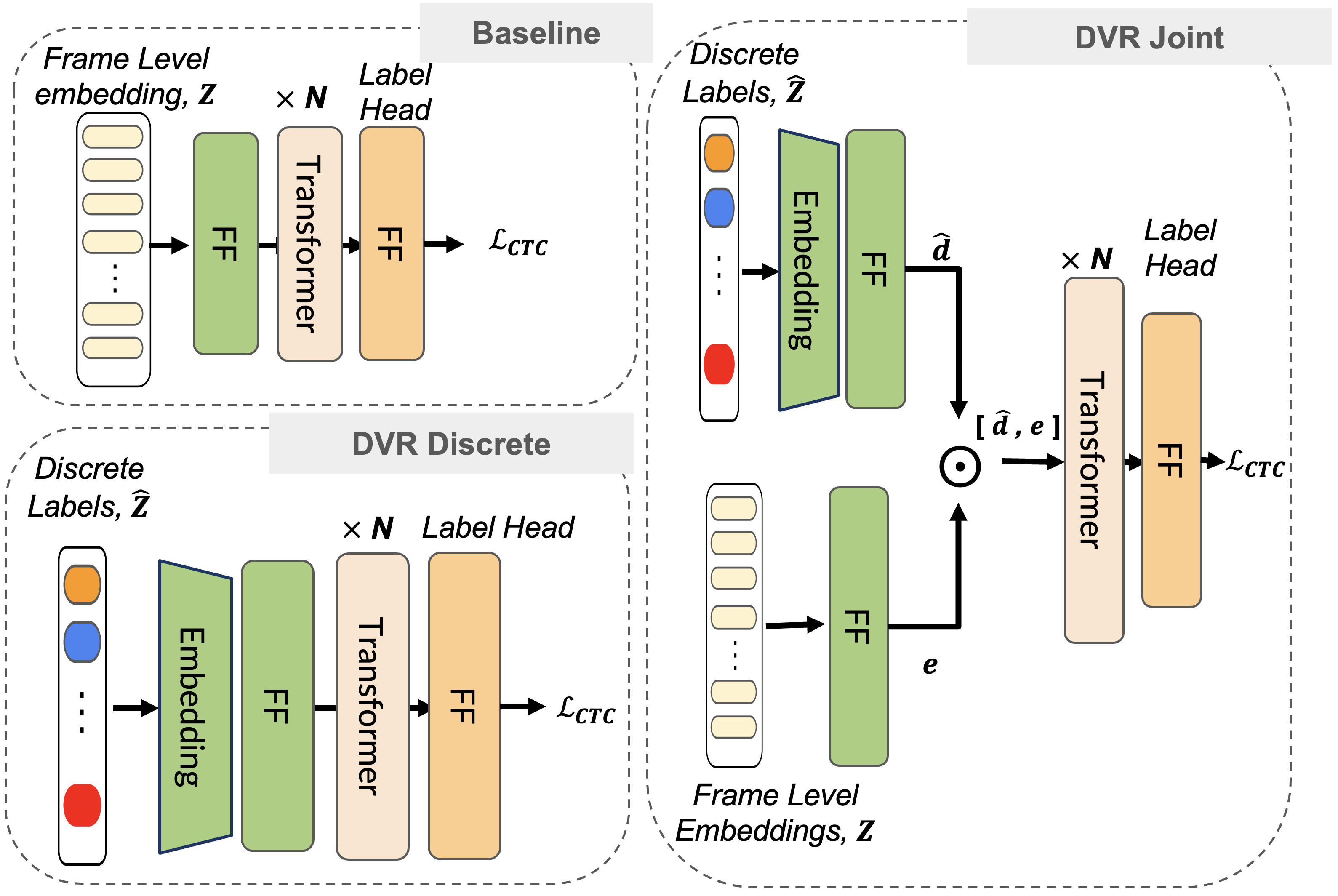}
\caption{\small{Baseline and DVR -- Discrete and Joint Model}}
\label{fig:model_architecture}
\end{figure*}

We opt for the large XLS-R
($1B$ parameters). Our preliminary analysis revealed limitation in the XLS-R in differentiating between acoustic sounds, such as \<د> \textipa{[d]}/ \<ض> \textipa{[d\super Q]} and \<ت> \textipa{[t]}/ \<ط> \textipa{[t\super Q]} present in MSA and DA. Consequently, we primed the model towards Arabic sounds by finetuning with 13 hours clean avaliable MSA data \cite{ardila2019common} for ASR task. We restricted the training to 5 epochs to prevent the risk of catastrophic forgetting of the pretrained representation \cite{goodfellow2013empirical}.

\subsection{Vector Quantization}

Vector Quantization \cite{makhoul1985vector, baevski2020wav2vec} is a widely used technique for approximating vectors or frame-level embeddings through a fixed codebook size. 
In our Vector Quantization (\textit{VQ}) modules (see Figure \ref{fig:exp_flow}), we pass forward a sequence of continuous feature vectors $Z = \{z_1, z_2, \ldots, z_T \}$ and then assign each $z_t$ to its nearest neighbor in the trained codebook, $\mathbb{C}_{k}$. In other words, each $z_t$ is replaced with the code $Q_{i}^{k} \in \mathbb{C}_{k}$ assigned to the centroid $G_{i}^{k}$ . The resultant discrete labels are quantized sequence $\hat{Z} = \{\hat{z}_1, \hat{z}_2, \ldots, \hat{z}_T\}$. These labels are expected to facilitate better proninciation learning and incorporate distinctive phonetic information in the subsequent layers.


\paragraph*{Training the Codebook}

For quantization, we utilized the k-means clustering model. We selected a random subset of frame-level representation for training the cluster model. Moreover, to select wide varieties of sound unit, we forced-aligned the available/automatic transcription of the datasets (see Section \ref{ssec:k-data}) with a GMM-HMM based ASR models. 
Using the timestamps, we then select SSL frame representations that aligned with wide varieties of sound labels.\footnote{$10k$ sample frames for each sound label.} We trained the codebook for different $k=\{128, 256, 512\}$

\subsection{Dialectal Units and Vowel Recognition (DVR) Model}

We explored two variants of DVR -- discrete and joint Model (as seen in Figure \ref{fig:model_architecture}). 
The discrete DVR takes only the discrete $\hat{Z}$ labels from the VQ as input, where as the joint module concatenate both the $\hat{Z}$ and $Z$ inside the subsequent layer. The resultant embeddings (for both model) are then passed to the transformer layers and the head feedforward layer. The DVR model is optimized with character recognition objective to identify arabic units.
%
\subsection{Baseline} 
As baseline, 
we used the frozen frame-level representation from the XLS-R model to pass to the feedforward layer followed by the transformers and output head. The architecture uses similar encoder as the DVR model (see Figure \ref{fig:model_architecture} Baseline). For brevity, we reported with the results of the second architecture (SSL frame-level representation with transformer-based encoder) as the baseline of the paper.

\section{ArabVoice15 Dataset}
\label{sec:data}

Spoken DA remains a low-resource language primarily due to the scarcity of transcription that can faithfully capture the diverse regional and borrowed sounds in the standard written format. Such lack of data posses significant challenge for speech and linguistic research and evaluation. In this study, we address this challenge by designing and developing ArabVoice15 test set. Furthermore, we have also enhanced a subset of the existing Arabic Commonvoice \cite{ardila2019common}, Ar:CV${_R}$ dataset with restored vowels, borrowed and dialectal sounds. In the following sections, we will discuss the datasets, preprocessing steps along with in detail annotation guidelines.


ArabVoice15 is a collection of 5 hours of speech utterances randomly selected from testset of ADI17 \cite{ali2019mgb} dataset, widely used for dialect identification task. For the ArabVoice15, we selected a total of 2500 utterance, $\approx 146 (\pm 3.6)$ utterance from each of the 15 Arab countries including: Algeria (ALG), Egypt (EGY), Iraq (IRA), Jordan (JOR), Saudi Arabia (KSA), Kuwait (KUW), Lebanon (LEB), Libya (LIB), Morocco (MOR), Palestine (PAL), Qatar (QAT), Sudan (SUD), Syria (SYR), United Arab Emirates (UAE), and Yemen (YEM). 
The average utterance duration: 7-8 seconds.
As for $Ar:CV{_R}$, we randomly extracted $21.38$ hours from the Ar:CV trainset, which we then mannually annotated at both verbatim and vowelized level (test $\approx$ 1hr).




\begin{table*}[!ht]
\scalebox{0.76}{
\begin{tabular}{l|cccc}
\toprule
\multicolumn{1}{c}{\textbf{Dataset}} & \textbf{Source of Data}                                                                                                                           & \textbf{Train (\#hrs)} & \textbf{Test (\#hrs)} & \textbf{Annotated with}                                                                                                                 \\ \hline\hline
\textit{Ar:CV${_R}$}$^+$                       & \begin{tabular}[c]{@{}c@{}}Subset from Arabic Common Voice \\   \cite{ardila2019common} Train split\end{tabular}                                                            & 1 hr ($^*$total 19 hrs)   & 1 hr                  & \begin{tabular}[c]{@{}c@{}}Restored short vowels, dialectal \\ and borrowed sounds\end{tabular}                                         \\\hline
\textit{AR:TTS-data}                    & \begin{tabular}[c]{@{}c@{}}Subset collected from available \\ test-to-speech speech corpus (2 speakers, one \\ from Egypt and Levantine region)\\ \cite{abdelali-2022-natiq,abdelali-2024-larabench,dalvi2023llmebench}\end{tabular} & 30 mins                & --                    & --                                                                                                                                      \\\hline
\textit{EgyAlj}                      & \begin{tabular}[c]{@{}c@{}}in-house, source Aljazeera Arabic channel, \\ containing MSA and Egy content\end{tabular}                                            & --                     & 1.8 hrs               & \begin{tabular}[c]{@{}c@{}}Semi-supervised transcription, \\ manually restored short vowels, \\dialectal and borrowed sounds.\end{tabular} \\ \hline\hline
\rowcolor{yellow!20}
\textit{ArabVoice15}$^+$                 & A small subset for ADI17 \cite{ali2019mgb} test set                                                                                                                 & --                     & 5 hrs                 & \begin{tabular}[c]{@{}c@{}}Transcribed with dialectal and \\ borrowed sound in consideration\end{tabular}              \\ \bottomrule                
\end{tabular}}
\caption{\small Train and Test dataset used for Dialectal Units and Vowel Recognition (DVR) model. $^*$ present total hours of data available and used to show the effect of training data size. $^+$ test data will be made available to the public.}
\label{tab:data}
\end{table*}


    
    



\paragraph*{Data Verbatim Pre-Processing}
We present a set of rules employed for data normalization, aiming to reduce annotators' tasks through a rule-based phonemic letter-to-sound approach in Arabic, as detailed in \cite{al2004phonetic}.
For vowelization, we initially applied diacritization (aka vowelization or vowel restoration) module present in the Farasa tool \cite{abdelali2016farasa}. We then applied the following rule-based phonemic letter-to-sound function to our dataset. This step also removed any Arabic letters that are not traditionally pronounced in spoken conversation.
\begin{itemize} [leftmargin=*, noitemsep,topsep=0pt,parsep=0pt,partopsep=0pt]
    \item \noindent For \<ا>   [\textipa{a\textlengthmark}] :
(i) If it appears within a word (not at the beginning) and is followed by two consonants, we delete it. For example, \<كتب الكتاب> [\textipa{ktb a\textlengthmark lktb}] becomes \<كتب لكتاب> [\textipa{ktb lktb}]. (ii) If it occurs at the beginning in the form of the definite article \<ال>, we replace it with [\textglotstop a]. For example, \<المعلم> [\textipa{a\textlengthmark lm\textrevglotstop lm/}] becomes \<ءَلمعلم> [\textipa{\textglotstop alm\textrevglotstop lm}].

\item For \<ل > [\textipa{l}] : We removed the Shamsi (Sun) [\textipa{l}], that refers to [\textipa{l}] in \<ال> followed by a Sun consonant\footnote{In Arabic grammar, there are two categories of letters: "sun letters" \<الحروف الشمسية> and "moon letters" \<الحروف القمرية>. These categories affect the pronunciation of the Arabic definite article \<ال> (al-). Sun letters are those Arabic letters that cause assimilation \<الإدغام> of the definite article \<ال> (al-) when they are prefixed to nouns, meaning the "l" sound of "al-" merges with the initial consonant of the noun. The assimilation occurs in pronunciation, but not in writing.

}
\<لنتثدذرزسشصضطظ>.
For example: \<الرحمان> [\textipa{a\textlengthmark lr\textcrh man}]  becomes \<ارحمان> [\textipa{a\textlengthmark r\textcrh man}]

\item For \<آ >, we replaced it wherever it occurred in the text with \<ءا> [\textglotstop a\textlengthmark].

\item For Hamza shapes (\<ء أ ؤ إ ئ>), we normalized them to \<ء> [\textglotstop].

\item For \<ا ى>, we normalized them to \<ا> [a\textlengthmark/].

\item For Tanwin diacritics (\< ً ٍ  ٌ> [\textipa{/un/}, \textipa{/in/}, \textipa{/an/}]) at the end of a phrase, we replaced it with a short vowel, and elsewhere, we turned it into \< َن>, \< ُن>, \< ِن> [\textipa{/un/}, \textipa{/in/}, \textipa{/an/}] to match the typical verbatim sounds.
\end{itemize}

\paragraph*{Annotation Guideline}
We gave extensive training to an expert transcriber, a native speaker from Egypt, to provide the written form for each word and its verbatim transcription. For example, if the word is \<قَلَم> [\textipa{qalam}] (pen), and the speaker said \<كَلَم>  [\textipa{kalam}], then the transcriber writes [\textipa{qalam/kalam}]. This is the summary of the annotation guidelines:


\begin{itemize} [leftmargin=*, noitemsep,topsep=0pt,parsep=0pt,partopsep=0pt]
\item For sounds that are not in MSA and have been borrowed from foreign languages, the following special letters\footnote{
The special letters used in the annotation process do not belong to the Arabic alphabet; instead, we borrowed them from Farsi sharing similar Arabic shapes, these letters were employed to represent distinct dialectal sounds.} are used:
\begin{itemize} [leftmargin=*, noitemsep,topsep=0pt,parsep=0pt,partopsep=0pt]
    \item \<چ> [\textipa{g}] as in the word \<جوجل> ``google'' which is written as \<جوجل> [\textipa{ju:jl}] / \<چوچل> [\textipa{gu:gl}].
    \item \<ڤ> [\textipa{v}] as in the word \<ڤيديو> ``video'' which is written as \<فيديو> [\textipa{fi:dyu:}] / \<ڤيديو> [\textipa{vi:dyu:}].
    \item \<پ> [\textipa{p}] as in the word \<إسبراي> ``spray'' which is written as \<سبراي> \ [\textipa{sbra:y}] / \<سپراي> [\textipa{spra:y}].
\end{itemize}




\item For dialectal sounds that are missed in MSA, the following special letters are used:

\begin{itemize} [leftmargin=*, noitemsep,topsep=0pt,parsep=0pt,partopsep=0pt]
    \item \<گ> (Gulf /Qaf/) as in the word \<عگال> which is written as \<عقال> / \<عگال>. 
    \item The Egyptian/Syrian/Lebanese \<ق> [\textipa{q}] is pronounced mostly as \<ء> [\textglotstop] as in \<قال>\ [qa:l] / \<ءال> [\textglotstop a:l].
    \item \<ڟ> (Egyptian/Lebanese /Z/) as in the word \<بيڟهر> is written as \<بيظهر> / \<بيڟهر>.
\end{itemize}
\end{itemize}


There are few words with special spellings that do not precisely reflect their pronunciation. In these cases, the transcriber writes both, as in the word \<هذا> \textipa[hadha] / \<هاذا> (/ha:dha/).
Numbers and some special symbols (ex: the percentage sign \%) are written in letters and are being judged according to speakers' pronunciation.

\noindent\textbf{Quality Control:} Detection of possible annotation errors was done automatically and doubtful cases were returned to the transcriber for review. In addition, a manual inspection of random sentences (10\%) from each file was performed. Any file below 90\% accuracy was returned for full correction.

\section{Experimental Design}




\subsection{Training Datasets and Resources}
\label{ssec:k-data}
\paragraph*{Datasets: Unspervised Codebook Generation} To train the codebook, we randomly selected utterances from publicly available resources. For Arabic sounds, we opt for utterances from official CommonVoice train set along with Arabic TTS data. Moreover, to add borrowed/special sounds missing in MSA phonetic set (e.g., /\textipa{g, v, p}/), we included publicly available English datasets like LibriSpeech \cite{panayotov2015librispeech}, and TIMIT \cite{garofolo1993darpa}.  
For the subsampling process, we opt for hybrid ASR systems\footnote{Trained on Arabic CommonVoice} for Arabic and Montreal Forced-Aligner\footnote{\url{https://github.com/MontrealCorpusTools/Montreal-Forced-Aligner.git}} for the English.
\paragraph*{Datasets: Spervised DVR Model}
To train the DVR model, we opt for a small training dataset to showcase our the efficacy of our proposed framework in low-resource setting. The details of dataset used for DVR is presented in Table \ref{tab:data}. For the training, we utilize dataset transcribed with restored vowels, borrowed and dialectal sounds. We used 1 hour 30 minutes of training data in this study.

\subsection{Model Training}

The Models, presented in Figure \ref{fig:model_architecture}, are optimized using Adam optimizer for 50 epochs with an early stopping criterion. The initial learning rate is $1 \times 10^{-4}$, and a batch size of $16$ is employed. The loss criterion is CTC loss, utilized for predicting verbatim sequences. 
The input dimension for the SSL frame-level representation is $d=1024$, the dimension of the discrete labels $d=k$.
For all the architectures in Figure \ref{fig:model_architecture}, the dimension of feedforward (FF) layer is 
$d=512$. For the DVR joint, the output from the FFs ($\hat{d}, e$) are concatenated to form $[\hat{d}, e]$ of dimension $d=1024$. These outputs are then passed to 2 transformer encoders each with 8 attention heads. Following, the encoded information is then projected to output head of dimension $V=39$ equivalent to the characters supported by the models. The total number of trainable parameters are Baseline:7.634M; DVR discrete:7.110M; and joint: 33.346M.

\subsection{Evaluation Measures}
We used \textbf{Davis-Bouldin index} (DBindex) to select the $k$ value for our codebook. The DBindex is widely used in clustering performance evaluation \cite{davies1979cluster}, and is characterized by the ratio of within-cluster scatter to between-cluster separation. A lower DBindex value is better, signifying compact clustering.
Following, we adapted the approach of \cite{hsu2021hubert} to evaluate the codebook quality using \textbf{Phone Purity}, \textbf{Cluster Purity}, and \textbf{Phone-Normalized Mutual Information} (PNMI). These measures use frame-level alignment of characters with discrete codes assigned to each frame. Phone purity measures the average frame-level phone accuracy, when we mapped the codes to its most likely phone (character) label. Cluster purity, indicates the conditional probability of a discrete code given the character label. PNMI measures the percentage of uncertainty about a character label eliminated after observing the code assigned. A higher PNMI indicates better quality of the codebook.
Moreover, we assessed the codebook quality by \textbf{human perception} tests as mentioned in the following section.
As for evaluating the dialectal sounds and short vowel recognition model, we reported Character Error Rate (CER) with and without restoring short vowels. 

\paragraph*{Human Perception Test Setup} We performed cluster quality analysis for $k=\{128, 256, 512\}$ following the steps of \cite{mao2018unsupervised, li2018unsupervised}. For our study, we defined each clusters (demoted by a code) as either Clean or Mix. Clusters are considered as Clean when 80\% of its instances are matched to one particular character, where as for Mix clusters, the instances are mapped to different characters.\footnote{Only characters above 20\% frequency are considered.} We hypothesise that the Mix clusters represent examples which can resembles closely to either two of canonical sound unit $/l1/$ and $/l2/$, or a mix of both $/l1\_l2/$.
We randomly selected 52 examples from each perceived Mix Clusters. We asked the four annotators (2 native and 2 non-native Arabic speakers) to categorize it into these four classes: more similar to $/l1/$, more similar to $/l2/$, a mix of both, or neither.

\section{Results and Discussion}
\label{sec:result}

\begin{figure*} [ht]
\centering
\includegraphics[width=1\textwidth]{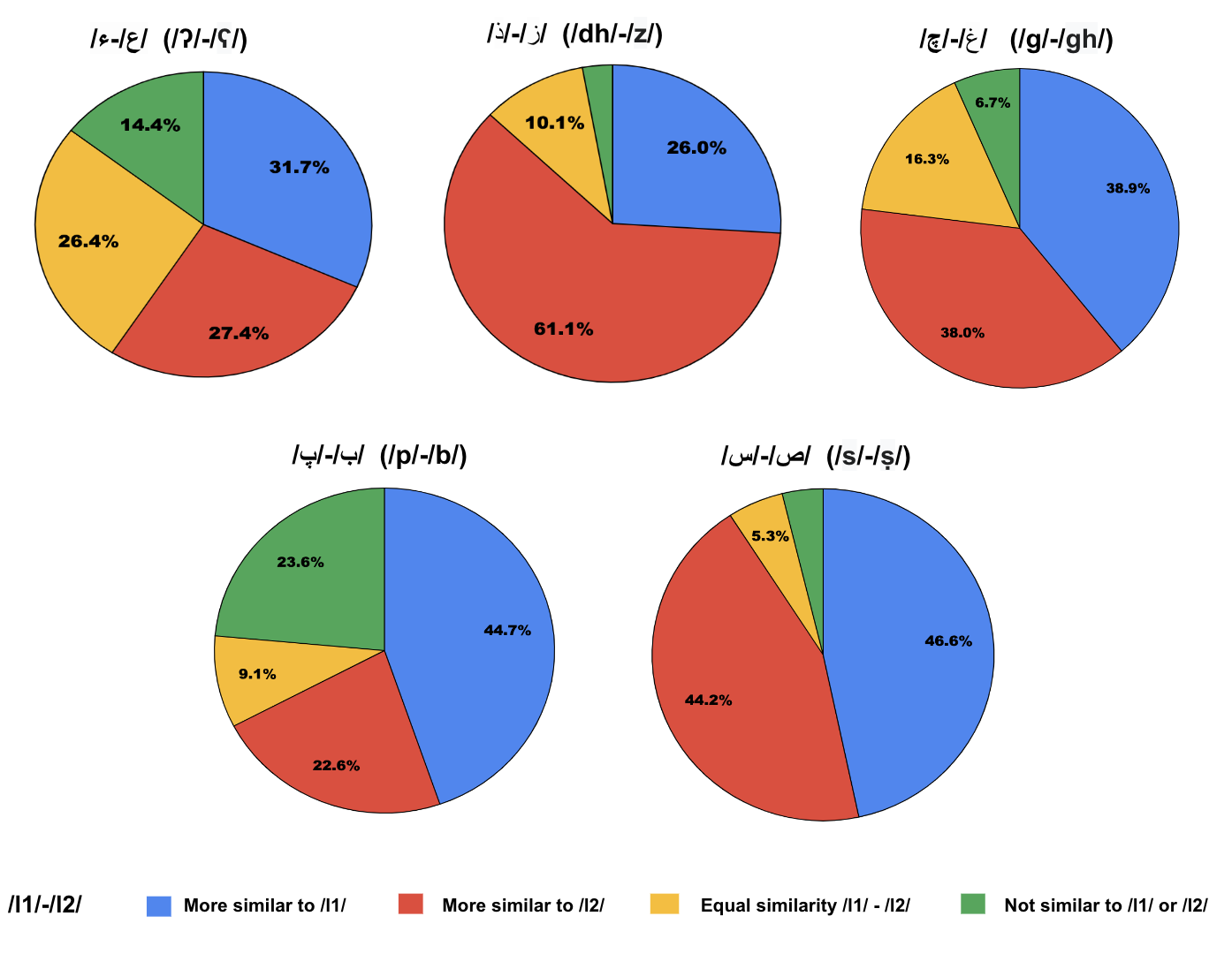}
\caption{\small The statistical results of perceptual tests of different sounds using cluster with $k=256$}
\label{fig:perceptual_testing}
\end{figure*}

\paragraph{Number of discrete codes in Codebook} We reported the DBindex for the codebook sizes $k=\{128, 256, 512\}$ in Table \ref{tab:purity}. We observed lower DBindex with $k=256$ indicating better codebook quality.
We further evaluated the codebook quality and reported purity measures with the Ar:CV${_R}$ testset only for brevity and CER with all the testsets. Our CER results shows the efficacy of the selected $k=256$ for most of the test sets. We observed that increasing codebook size improves the purity and the PNMI. We noticed, the gain in cluster stability between $k=256$ {\em vs} $k=516$ is not very large with respect to the performance and computational cost.
Hence we selected the codebook $\mathbb{C}$ of size $k=256$ for all the experiments. 

\begin{table}[!ht]
\centering
\scalebox{0.9}{
\begin{tabular}{c|c|c|c}
\toprule
\multicolumn{1}{c|}{\textit{k}} & \multicolumn{1}{c|}{\textit{128}} & \multicolumn{1}{c|}{ \underline{\textit{256}}} & \multicolumn{1}{c}{ \textit{512}} \\ \midrule\midrule
\rowcolor{yellow!20}
\multicolumn{4}{c}{$\mathbb{C}$ size $k$ selection criterion} \\ \hline\hline
\multicolumn{1}{l|}{DBindex ($\downarrow$)} & { 2.59} & { \textbf{2.57}} & { 2.7} \\ \hline\hline
\rowcolor{yellow!20}
\multicolumn{4}{c}{Purity Measures: Ar:CV$_R$ testset} \\ \hline
\multicolumn{1}{l|}{Phone Purity ($\uparrow$)} & { 0.600} & { 0.641} & { 0.672} \\ \hline
\multicolumn{1}{l|}{Discrete Code Purity ($\downarrow$)} & 0.436  &  0.289 & 0.236 \\ \hline
\multicolumn{1}{l|}{PNMI ($\uparrow$)} & { 0.343} & { 0.418} & { 0.495} \\ \hline\hline
\rowcolor{yellow!20}
\multicolumn{4}{c}{CER ($\downarrow$): Borrowed and Dialectal Unit Recognition } \\ \hline\hline
\multicolumn{1}{l|}{Ar:CV${_R}$} & \multicolumn{1}{c|}{0.149} & \multicolumn{1}{c|}{0.108} & \multicolumn{1}{c}{0.107} \\
\multicolumn{1}{l|}{EgyAlj} & \multicolumn{1}{c|}{0.246} & \multicolumn{1}{c|}{0.206} & \multicolumn{1}{c}{0.218} \\
\multicolumn{1}{l|}{ArabVoice15} & \multicolumn{1}{c|}{0.465} & \multicolumn{1}{c|}{0.447} & \multicolumn{1}{c}{0.462} \\ \hline
\multicolumn{1}{l|}{\textbf{Average}} & \multicolumn{1}{c|}{0.287} & \multicolumn{1}{c|}{\textbf{0.254}} & \multicolumn{1}{c}{0.262} \\\bottomrule
\end{tabular}}
\caption{\small Quality evaluation of discrete codes based on DBindex, purity measures and CER for 3 test sets. }
\label{tab:purity}
\end{table}

\begin{table}[!ht]
\centering
\scalebox{1}{
\begin{tabular}{l|ccc}
\toprule
\multicolumn{1}{c}{CER} & $Z$   & $D_D$   & $D_J$ \\
\midrule\midrule
\rowcolor{yellow!20}
\multicolumn{4}{c}{Training Data}                                                                          \\ \hline\hline
\multicolumn{4}{c}{1hr 30min}                                                                          \\ \hline\hline
Ar:CV${_R}$                     & 0.113                 & 0.108                    & \textbf{0.094}         \\
EgyAlj                  & 0.252                 & \textbf{0.206}           & 0.231                   \\
AraVoice15              & 0.536                 & \textbf{0.447}           & 0.464                  \\ \hline

\multicolumn{4}{c}{3hrs 30min}                                                                          \\\hline\hline
Ar:CV${_R}$                     & 0.103                 & 0.108                    & \textbf{0.096} \\
EgyAlj                  & 0.270                 & \textbf{0.241}           & 0.253     \\
AraVoice15              & 0.497                 & \textbf{0.470}           & 0.483   \\ \bottomrule    

\rowcolor{gray!20}
\multicolumn{4}{c}{5hr 30min} \\ \hline\hline
Ar:CV${_R}$  & \textbf{0.095}        & 0.110                    & 0.099 \\
EgyAlj & 0.257 & \textbf{0.245} & 0.248 \\
AraVoice15 & 0.485 & \textbf{0.477} & 0.491 \\ \hline\hline

\multicolumn{4}{c}{$\sim$20 hrs} \\\hline\hline
Ar:CV${_R}$ & \textbf{0.099}        & 0.108                    & 0.101 \\
EgyAlj & 0.264   & 0.244      & \textbf{0.227} \\
AraVoice15 & 0.492                 & 0.478                    & \textbf{0.457} \\ \bottomrule

\end{tabular}}
\caption{\small Reported CER performance for borrowed and dialectal unit recognition task with Baseline ($Z$), DVR Discrete ($D_D$) and DVR Joint ($D_J$) models, for all three test sets and different training data sizes.  }
\label{tab:cer}
\end{table}

\begin{table}[!ht]
\centering
\scalebox{1}{
\begin{tabular}{l|cccc}
\toprule
\multicolumn{1}{c|}{CER}     & Farasa      & $Z$   & $D_D$   & $D_J$   \\ \midrule\midrule
Ar:CV${_R}$        & 0.279    & 0.123      & 0.278          & \textbf{0.118}       \\
EgyAlj        & \textbf{0.250} & 0.279      & 0.395          & 0.274       \\\bottomrule
\end{tabular}}
\caption{\small Reported CER for  Farasa, Baseline ($Z$), DVR Discrete ($D_D$) and DVR Joint ($D_J$) models for two test sets. Training set of 1 hour 30 minutes.}
\label{tab:shortV}
\end{table}

\paragraph{Perceptual test of Codebook}
We averaged annotator judgments across four categories for all Mix clusters, revealing no clear majority and highlighting the listeners' difficulty in categorically labeling audio within these clusters.
In aligned with \citet{mao2018unsupervised, li2018unsupervised}, we also conclude that these mixed labels genuinely exist and cannot be precisely characterized by any conventional given label. We present some of our findings of the perceptual test in Figure \ref{fig:perceptual_testing} for 5 different Mix clusters with average judgment per category.



\paragraph{Dialectal Unit Recognition Performance}
We reported the performance of the proposed DVR discrete and joint model in Table \ref{tab:cer} for borrowed and dialectal unit recognition task. Our results shows the efficacy of the DVR models over the baseline specially for dialectal test sets (ArabVoice and EgyAlj). We observed for borrowed and dialectal unit recognition, the discrete model outperforms the joint model significantly. Breakdown of the performance for 15 countries are presented in Appendix \ref{Appendix2}. 

\paragraph{Impact of Training Data size} Table \ref{tab:cer} also shows the impact of the training data size. We observed for dialectal unit recognition, our DVR discrete model outperforms the other two models significantly with limited data sets of $\{1hr30min, 3hr30min, 5hr30min\}$. We see an improvement in performance from 1hr30min to 3hr30min settings. However, beyond a certain data threshold, the improvements plateaued. 

\paragraph{Performance for short vowel restoration} 
For short vowel restoration (in Table \ref{tab:shortV}), we observed that the added frame-level embeddings (in DVR joint) improve the recognition performance. We also observed that the baseline model performs comparably with DVR joint. This indicates that the restoration of short vowels benefits from high dimensional fine-grained information compare to using few discrete codes. We also compared the CER with Farasa -- state-of-the-art text-based dicretization tool \cite{abdelali2016farasa}. We observed the acoustic models outperform Farasa by a large margin, especially for common voice subset. However, Farasa excelled in formal content -- news content presented in EgyAlj testset.


\section{Conclusion}
In this study, we propose a novel dialectal sound and short vowel recovery framework that utilizes a handful of discrete codes to represent the variability in dialectal Arabic. We also observed with only 256 discrete labels, the borrowed and dialectal sound recognition model outperforms both baseline and joint (discrete code with frame-level SSL representation) models by $\approx 7\%$ CER improvement. For restoring vowels, we noticed SSL embeddings play a bigger role. 
Our findings indicate the efficacy of the discrete model with small training datasets. To foster further research in dialectal Arabic, we introduced, benchmarked, and released ArabVoice15 -- a dialectal verbatim transcription dataset containing utterances from 15 Arab countries. In the future, we will apply the framework to more dialects and other dialectal languages.


\label{sec:bibtex}
\section*{Limitations}
The diversity of representation and the size of ArabVoice15 could limit the conclusion to generalize in all Arabic dialects due to variability in dialectal sounds. Although the annotator was an expert transcriber and received extensive training, their dialect may have led to some bias in judgment. 

\section*{Ethics Statement}
For the research work presented in this paper on the Dialectal Sound and Vowelization Recovery (DSVR) framework, we have adhered to the highest ethical standards. All the speech/audio data used in this study were already publicly available. 
The human perception tests for our evaluation process were designed with a commitment to fairness, inclusivity, and transparency. The participants were selected keeping in mind balancing gender and nativity. Listeners were fully briefed on the nature of the research and their rights as participants, including the right to withdraw at any time without consequence. However as we mentioned in the limitation section, we cannot guarantee any human bias toward any dialectal sound or preference.

\bibliography{anthology,custom}
\bibliographystyle{acl_natbib}

\appendix

\section{Appendix}
\label{sec:appendix}

\begin{figure*} [ht]
\centering
\includegraphics[width=1.0\textwidth]{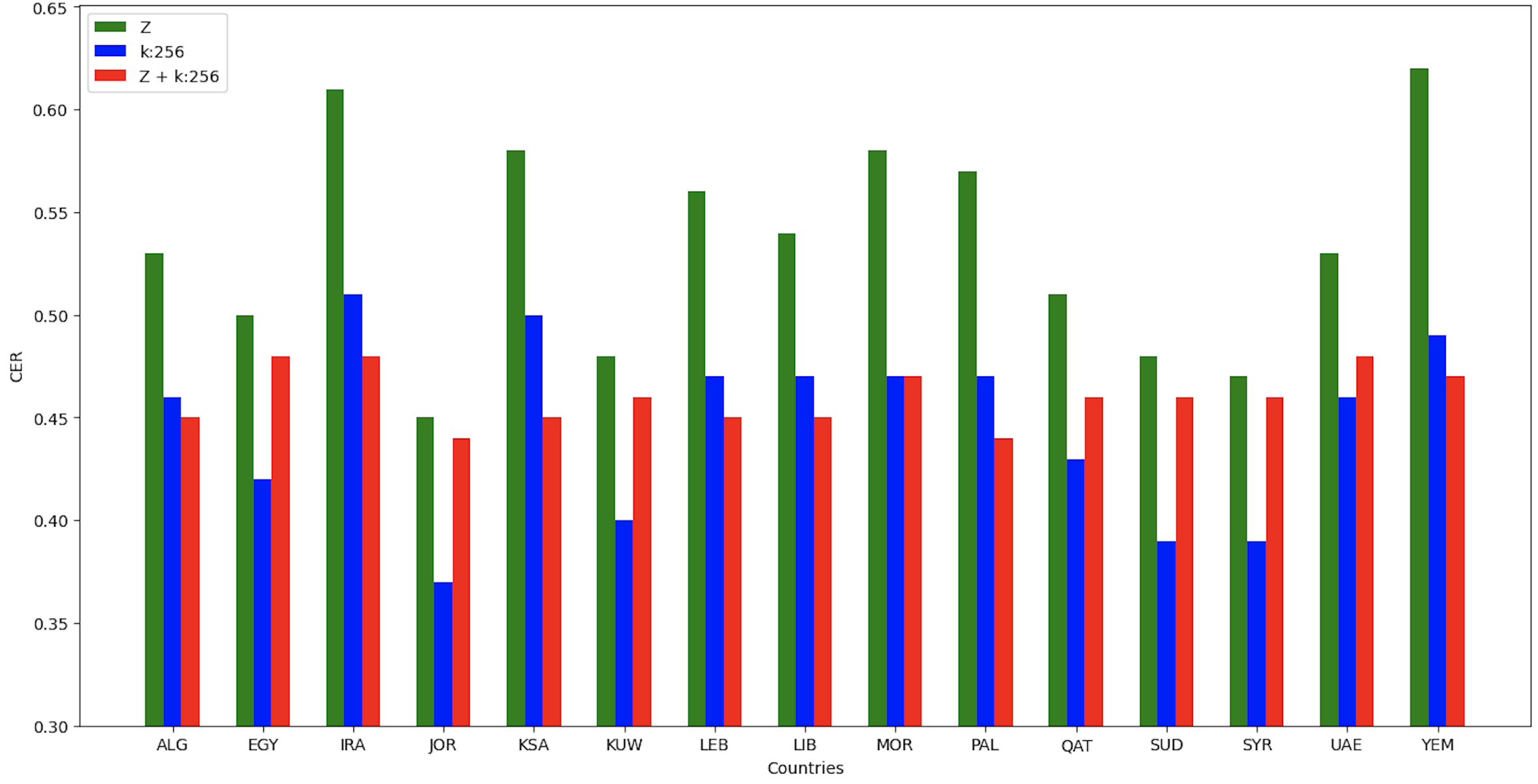}
\caption{Reported CER for test utterances from 15 Arab countries for three models Baseline (Z), DVR discrete (k:256) and DVR joint (Z+k:256)}
\label{fig:barplot}
\end{figure*}

\subsection{Sound Analysis}
\label{Appendix1}

\begin{figure*} [ht]
\centering
\includegraphics[width=1.0\textwidth]{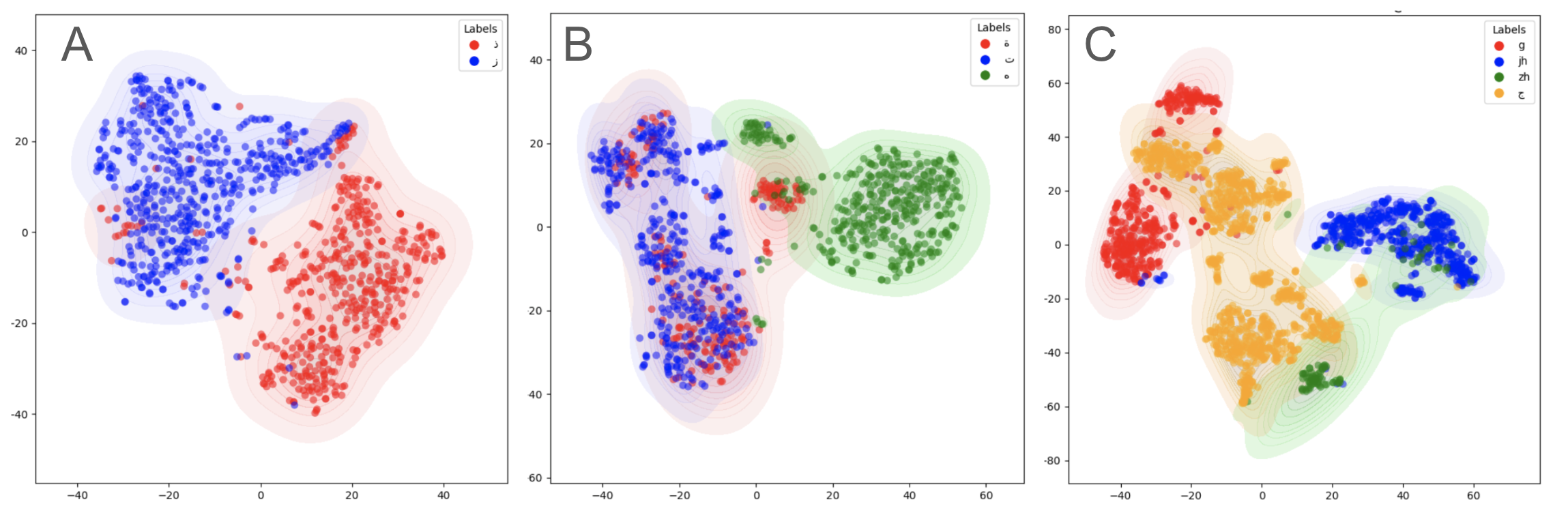}
\caption{2D t-SNE Projection of Frame-Level Presentations Extracted Randomly from Finetuned Arabic XLS-R. A. Pairs (\<ز ذ>) \textipa[\dh \ z]. B. Sounds (\<ت ة ه>)  \textipa[h t]. C. Pairs (\<ج>     [\textipa{d\textctz}], zh [\textctz], g 
).}
\label{fig:dialects_samples}
\end{figure*}

In Figure \ref{fig:dialects_samples}, we have depicted potential confusion between specific sounds in MSA and Arabic dialects. Utilizing a Hidden Markov Model-Time Delay Neural Network (HMM-TDNN) model\footnote{\url{https://kaldi-asr.org/models/m13}}, trained with MGB-2 \cite{ali2016mgb} for Arabic, we aligned randomly selected samples from the original datasets of CommonVoice Arabic and EgyAlj. For the English dataset TIMIT, we used the provided ground truth alignment.

After aligning speech signals with their original unvowelized character-based transcriptions, we matched frame-level features extracted from XLS-R (see Section \ref{sec:xlsr}) with their corresponding characters. In Figure \ref{fig:dialects_samples}.A, we randomly selected $1000$ samples associated with \<ز> [\textipa{z}] and $1000$ samples associated with \<ذ> [\textipa{\dh}] from CommonVoice Arabic. Despite CommonVoice Arabic being considered as clean MSA speech data with good pronunciation, we observed that some samples of \<ذ> [\textipa{\dh}] were clustered with \<ز> [\textipa{z}], primarily explained by the speakers getting influenced by their dialectal variations, as discussed in Section \ref{sec:arabsounds}.

Figure \ref{fig:dialects_samples}.B displays the selection of three characters: \<ت> [\textipa{t}],
\<ة> [\textipa{t}/\textipa{h}],
\<ه> [\textipa{h}]. Notably, \<ة> is at times pronounced as [\textipa{t}] and at other times as [\textipa{h}]. Although rule-based methods \cite{halabi2016phonetic} can predict when it will correspond to which sound, applying these rules in everyday spoken language, where people don't follow rule based pronunciation, proves challenging. The figure reveals two main clusters for [\textipa{t}] and [\textipa{h}], with vectors associated with \<ة> scattered between these clusters, highlighting the aforementioned point.

Figure \ref{fig:dialects_samples}.C illustrates the selection of four labels: Arabic \<ج> [ [\textipa{d\textctz}],
and English phonemes (zh, g, jh) [\textctz, \textipa{g}, \textipa{d\textctz}]. We selected $1000$ Arabic samples of \<ج> from CommonVoice Arabic and EgyAlj, along with $500$ samples for each of the English phonemes. It became apparent that the Arabic sound \<ج> is distributed across different English pronunciations (zh, g, and jh), indicating dialectal variations in the pronunciation of \<ج>.

\subsection{Country-wise DVR performance}
\label{Appendix2}

In this section, we present the aforementioned results discussed in Section \ref{sec:result}. Figure \ref{fig:barplot} displays CER results for the Baseline (Z), SVR Discrete (k:256), and DVR joint (Z+k:256) models trained on 1H30min of data, tested on AraVoice15. We analyze the CER results for each dialect individually. Our observations reveal that SVR Discrete (k:256) and DVR joint (Z+k:256) consistently outperform the Baseline (Z) across all dialects, exhibiting a substantial performance gap in MOR, YEM, PAL, and IRA dialects. Moreover, SVR Discrete (k:256) and DVR joint (Z+k:256) exhibit similar performance across the majority of the 15 dialects (10/15), with notable disparities observed in JOR, SUD, SYR, where a discernible performance gap is evident.


\end{document}